\documentclass[english, twocolumn, 10pt, aps, superscriptaddress, floatfix, showpacs, prb, citeautoscript]{revtex4-1}
\pdfoutput=1
\usepackage[utf8]{inputenc}
\usepackage[T1]{fontenc}
\usepackage{verbatim}
\usepackage{units}
\usepackage{mathtools}
\usepackage{amsmath}
\usepackage{amssymb}
\usepackage{graphicx}
\usepackage{wasysym}
\usepackage{layouts}
\usepackage{siunitx}
\usepackage{bm}
\usepackage{xcolor}
\usepackage[colorlinks, citecolor={blue!50!black}, urlcolor={blue!50!black}, linkcolor={red!50!black}]{hyperref}
\usepackage{bookmark}
\usepackage{tabularx}
\usepackage{microtype}
\usepackage{babel}
\usepackage{float}

\renewcommand{\comment}[2]{#2}

\DeclarePairedDelimiter\abs{\lvert}{\rvert}
\DeclarePairedDelimiter\norm{\lVert}{\rVert}

\makeatletter
\let\oldabs\abs
\def\abs{\@ifstar{\oldabs}{\oldabs*}}
\let\oldnorm\norm
\def\norm{\@ifstar{\oldnorm}{\oldnorm*}}
\makeatother

\newcolumntype{L}[1]{>{\raggedright\arraybackslash}p{#1}}
\newcolumntype{C}[1]{>{\centering\arraybackslash}p{#1}}
\newcolumntype{R}[1]{>{\raggedleft\arraybackslash}p{#1}}

\begin{document}
\title{Geometric focusing of supercurrent in hourglass-shaped ballistic Josephson junctions}
\author{Muhammad Irfan}
\affiliation{Kavli Institute of Nanoscience, Delft University of Technology, P.O. Box 4056, 2600 GA Delft, The Netherlands}
\affiliation{Department of Physics and Applied Mathematics, Pakistan Institute of Engineering and Applied Sciences, Nilore, Islamabad, 45650, Pakistan}
\author{Anton R. Akhmerov}
\affiliation{Kavli Institute of Nanoscience, Delft University of Technology, P.O. Box 4056, 2600 GA Delft, The Netherlands}

\begin{abstract}
The response of superconductor-normal-metal-superconductor junctions to magnetic field is complicated and non-universal because all trajectories contributing to supercurrent have a different effective area, and therefore acquire arbitrary magnetic phases.
We design a hourglass-shaped Josephson junction where due to the junction symmetry the magnetic phase of every trajectory is approximately equal.
By doing so we are able to increase a critical field of the Josephson junction to many flux quanta per junction area.
We then analyse how breaking the symmetry condition increases the sensitivity of the junction, and show that our device allows to detect supercurrent carried by ballistic trajectories of Andreev quasiparticles.
\end{abstract}

\maketitle

\section{Introduction}

\comment{Fraunhofer pattern in Josephson junctions is the magnetic oscillation of supercurrent due to the quantum interference of different trajectories.}
The Fraunhofer pattern~\cite{rowell_magnetic_1963} is a macroscopic quantum interference phenomenon in Josephson Junctions where critical current oscillates in response to an applied magnetic field in a fashion similar to the Fraunhofer diffraction of light passing through a single slit.
The applied magnetic field spatially modulates the phase which a quasiparticle acquires while traversing from one superconductor to another.
Because the contribution of each trajectory to supercurrent is an oscillatory function of this phase, contributions of different trajectories interfere.
Because of being able to distinguish different trajectories, Fraunhofer measurements are used to determine a spatial distribution of supercurrents~\cite{dynes_supercurrent_1971,hui_proximity-induced_2014,hart_induced_2014, pribiag_edge-mode_2015,allen_spatially_2016}.
Importantly, such measurements allow to distinguish current carried by the edge states from bulk conduction.

\begin{figure}[tbh]
\includegraphics[width=0.98\linewidth]{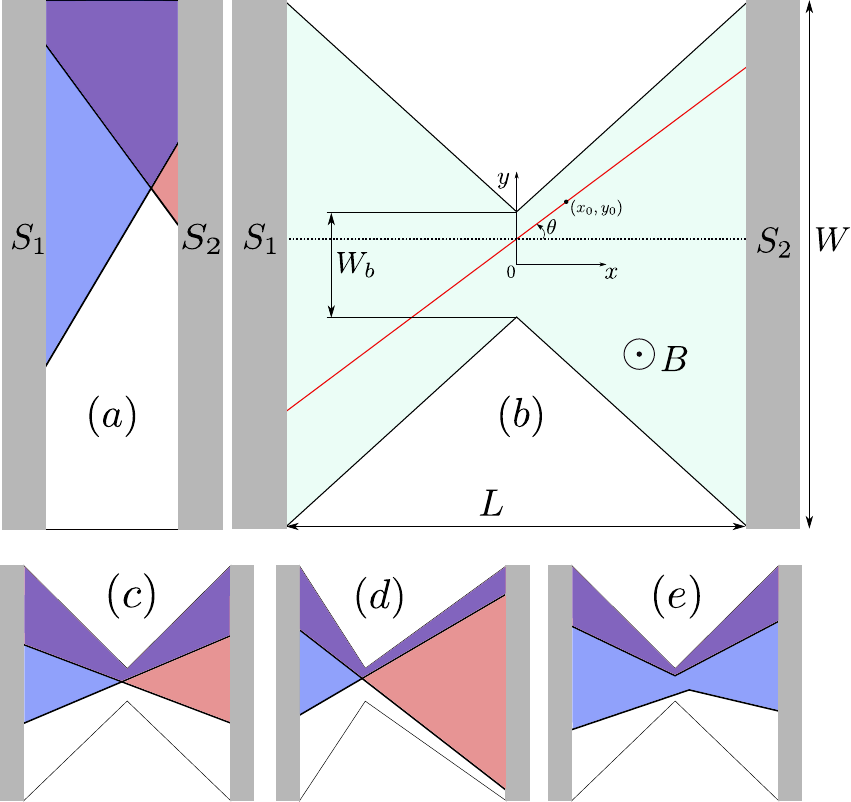}
\caption{
\label{Schematic}
Panel (a): Two Andreev trajectories (black lines) in an SNS junction accumulate a magnetic phase proportional to the area enclosed by such a trajectory (shaded regions).
Panel (b): In an hourglass-shaped SNS junction with a narrow opening $W_b$ all current-carrying trajectories pass through the middle.
Panel (c): The magnetic phases acquired by these trajectories in a symmetric device are approximately equal.
Breaking the reflection symmetry [panel (d)] or introducing disorder scattering [panel (e)] makes the magnetic phases different.}
\end{figure}

\comment{In a ballistic SNS JJ, different trajectories acquire different phases and the identification of their ballistic nature from Fraunhofer measurement is difficult.}
In a ballistic superconductor-normal-metal-superconductor (SNS) Josephson junction (JJ), different Andreev trajectories acquire different phases depending upon the path they follow.
The acquired phase is proportional to the trajectory area, as illustrated in Fig.~\ref{Schematic}(a).
The Fraunhofer pattern due to the interference of these trajectories depend on the geometry of the device \cite{barzykin_coherent_1999, cuevas_magnetic_2007, chiodi_geometry-related_2012, alidoust_-state_2013, amado_electrostatic_2013}.
Furthermore, Hendrik et. al., \cite{meier_edge_2016} show that the Fraunhofer pattern is sensitive to the reflection from the edges of the device.
At low magnetic field, the edge effects make the critical current nonzero at all values of the magnetic field and on the other accelerate the overall suppression of the critical current.
Both of these effects do not require ballistic trajectories.
It is therefore hard to identify the ballistic nature of Andreev trajectories from a Fraunhofer measurement, and in particular Refs.~\onlinecite{dynes_supercurrent_1971,hui_proximity-induced_2014} present a universal algorithm for interpreting any dependence of critical current on magnetic field as an inhomogeneous tunnel junction.

\comment{For a SNS geometry of hourglass shape, all the trajectories accumulate approximately same phase, which makes this geometry useful to probe these trajectories.}
Here, we design a device allowing to detect ballistic supercurrent based on a \emph{qualitative} change in the Fraunhofer pattern.
We show that in an hourglass-shaped JJ, shown in Fig.~\ref{Schematic}(b), the trajectories approximately accumulate the same phase, as shown in Fig.~\ref{Schematic}(c).
This phase matching condition provides a constructive interference of supercurrent also at high magnetic fields and results in a slow decay of critical current with magnetic field.
Breaking the spatial symmetry, by making the device geometrically asymmetric (Fig.~\ref{Schematic}(d)), by disorder (Fig.~\ref{Schematic}(e)), or by applying an asymmetric gate potential then restores the conventional Fraunhofer pattern.

\comment{Organization of the paper.}
The organization of the paper is as follows. We first introduce the physical system in Sec.~\ref{sec:system}. In Sec.~\ref{sec:quasiclassic}, we present the quasiclassical analysis of supercurrent in an hourglass device. In Sec.~\ref{sec:scattering-formalism}, we introduce the scattering matrix formalism and support our conclusions using numerical simulations based on a quantum-mechanical model. Finally, we summarize our analysis in Sec.~\ref{sec:discussion}.

\section{System}
\label{sec:system}
\comment{We consider an hourglass shape SNS JJ with a perpendicular magnetic field normal to the plane of JJ.}
We consider an hourglass-shaped Josephson junction with the separation between the superconducting contacts $L$, contact width $W$, and the bottleneck width $W_b$ as shown in Fig.~\ref{Schematic}(b).
The magnetic field $B$ in the scattering region is constant and perpendicular to the junction plane while being completely expelled from the superconductors.
We choose the Landau gauge, resulting in the the vector potential $\bm{A} = (-By \hat{x}, 0)$.
The Hamiltonian of the scattering region reads:
\begin{equation}
\label{eq:H-2DEG}
H = \frac{\left(\bm{p} - e\bm{A}\right)^2}{2m} - \mu +V(\bm{r}),
\end{equation}
with $\bm{p}$ the momentum operator, $e$ the electron charge, $\mu$ the chemical potential, $m$ the quasiparticle mass, and $V(\bm{r})$ the electrostatic potential in the scattering region.
While modeling superconducting leads, we assume a step-like superconducting pairing potential
\begin{eqnarray}
\Delta = \begin{cases}
    \Delta e^{i\phi_L} & {x<-L/2},\\
    \Delta e^{i\phi_R} & {x > L/2}, \\
    0 & {-L/2 \leq x\leq L/2},
  \end{cases}
\end{eqnarray}
with $\phi_L$ ($\phi_R$) the superconducting phase in left (right) lead.

\section{Quasiclasscial calculation of supercurrent}
\label{sec:quasiclassic}

\comment{We calculate supercurrent using quasiclassical trajectory + short junction approach.}
We start with a quasiclassical trajectory approach following Ref.~\onlinecite{ostroukh_two-dimensional_2016} to calculate supercurrent through the JJ before turning to a quantum-mechanical treatment.
The main underlying assumption  for quasiclassics is that the Fermi wavelength is much smaller than any feature of the system geometry.
Additionally, we consider the low field regime where the cyclotron radius is much larger than the system size, and trajectories are composed of segments of straight lines.
Supercurrent is then carried by closed trajectories where an electron originates from one superconductor, reaches another one, transforms into a hole via Andreev reflection, retraces back its original path, and finally transforms back into an electron via another Andreev reflection.
We focus on the short junction limit ($\hbar v_F / L \gg \Delta$ with $v_F$ the Fermi velocity), where every such trajectory supports a bound state with an energy
\begin{equation}
E =\pm \Delta \cos\left(\phi/2 - \xi/2\right).
\end{equation}
Here $\phi$ is the superconducting phase difference between the two terminals, and $\xi$ is the path-dependent magnetic phase:
\begin{equation}
\xi = \frac{2e}{\hbar} \int\limits_{S_{1}} ^{S_{2}} \bm{A}d\bm{l},
\label{eq:P-phase}
\end{equation}
with $e$ the electron charge and $\hbar$ the reduced Planck’s constant.
Assuming the short junction limit simplifies the numerical simulations, however we expect that relaxing this approximation will not alter our conclusions.

The supercurrent due to a single trajectory at a temperature $T$ reads:
\begin{align}
\label{eq:supcur}
\delta I &= -\frac{2e}{\hbar} \tanh\left(E/2k_BT\right)\frac{dE}{d\phi} \\
 &=\frac{e \Delta}{\hbar}\sin\left(\phi/2-\xi/2\right) \tanh\left[\frac{\Delta \cos\left(\phi/2-\xi/2\right)}{2k_BT}\right],\nonumber
\end{align}
with $k_B$ the Boltzmann's constant.
We further simplify the above expression by assuming sufficiently large temperature $k_BT \approx \Delta$ which leads to:
\begin{equation}
\delta I \approx \frac{e \Delta^2}{4\hbar k_BT}\sin\left(\phi-\xi\right).
\end{equation}
Because of the device geometry, most supercurrent-carrying trajectories do not scatter of the sample boundaries.
Therefore in our gauge choice, the phase $\xi$ for a trajectory passing through a point $(x_0, y_0)$ and making an angle $\theta$ with the $x$-axis is:
\begin{equation}
\xi = \frac{2eBL}{\hbar}\left(y_0 - x_0 \tan\theta\right).
\end{equation}
We calculate the total supercurrent through the junction by integrating over all the possible trajectories.
The integral simplifies upon setting $x_0=-L/2$:
\begin{equation}
I = \frac{k_F}{2\pi} \int\limits_{-W/2}^{W/2} dy_0 \int\limits_{\theta_{\min}}^{\theta_{\max}}\delta I \left(-L/2, y_0, \theta\right)\cos\theta d\theta.
\label{eq:supercurrent-symmetric}
\end{equation}
The requirement that a trajectory does not reflect at a boundary reads:
\begin{subequations}
  \label{eq:theta}
  \begin{align}
    &\theta_{\min} < |\theta| < \theta_{\max},\\
    &\theta_{\min} = \arctan\left[\max\left(\frac{-W_b - 2y_0}{L}, \frac{-W/2 - y_0}{L}\right)\right],\\
    &\theta_{\max} = \arctan\left[\min\left(\frac{W_b - 2y_0}{L}, \frac{W/2 - y_0}{L}\right)\right].
  \end{align}
\end{subequations}

\comment{We also consider the case of an asymmetric hourglass JJ and calculate the expression for supercurrent.}
In an asymmetric hourglass junction the bottleneck position is shifted by an offset $\delta L$ towards one of the superconducting leads, such that position of the bottleneck is at a distance $L_1 = L/2-\delta L$ from one lead and $L_2 = L/2+\delta L$ from the other.
If the offset $\delta L > W_b/2$ then straight trajectories starting from the top or bottom corners of the left superconducting lead do not reach the other lead.
This results in the modification of the integration limits in Eq.~\eqref{eq:supercurrent-symmetric} from $W$ to the effective junction width
\begin{equation}
W_{eff} = \frac{L-2\delta L}{L+2\delta L}\left(W/2+W_b/2\right)+W_b/2.
\end{equation}
The limits of the integral over angle $\theta_\textrm{min}$ and $\theta_\textrm{max}$ change to
\begin{subequations}
  \label{eq:theta-asym}
  \begin{align}
    &\theta_{\min} < |\theta| < \theta_{\max},\\
    &\theta_{\min} = \arctan\left[\max\left(\frac{-W_b/2 - y_0}{L_1}, \frac{-W/2 - y_0}{L}\right)\right],\\
    &\theta_{\max} = \arctan\left[\min\left(\frac{W_b/2 - y_0}{L_1}, \frac{W/2 - y_0}{L}\right)\right].
  \end{align}
\end{subequations}

\begin{figure}[tbh]
\includegraphics[width=0.5\textwidth]{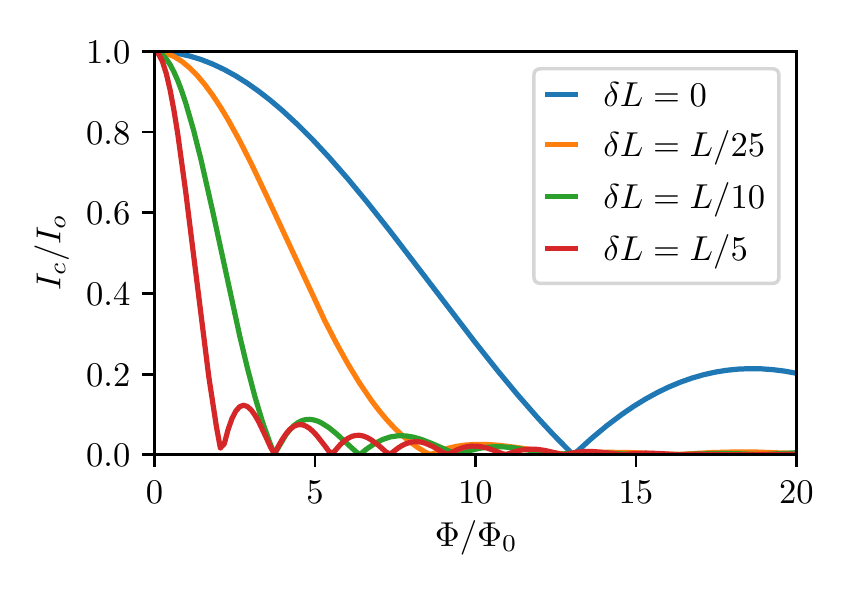}
\caption{
\label{QuasiclassicsFP}
Critical current as a function of magnetic flux $\Phi = B (W_b+W)L/2$ through the normal scattering region, calculated quasiclassically for symmetric and asymmetric hourglass-shaped Josephson junctions of dimension $L=W$ and $W_b = L/25$.
The asymmetry is controlled by $\delta L$, the displacement of the hourglass bottleneck from the middle of the device along $x$-axis.}
\end{figure}

\comment{The comparison of Fraunhofer pattern for symmetric and asymmetric JJ shows that the higher asymmetry leads to a regular pattern as opposed to the symmetric JJ.}
In Fig.~\ref{QuasiclassicsFP}, we show the Fraunhofer patterns for a symmetric and three different asymmetric JJs as a function of magnetic flux $\Phi = B (W_b+W)L/2$ through the device.
We confirm our expectation that the critical current for the symmetric hourglass device is less sensitive to the magnetic field and decays slower than that in a regular Fraunhofer pattern.
Specifically, the critical current vanishes at a magnetic field scale $B^\ast \sim \Phi_0/W_b L$.
Making the device asymmetric increases the sensitivity of the supercurrent to magnetic field, making it characteristic field scale $B^\ast \sim \Phi_0/WL$, similar to a conventional SNS junction.

\begin{figure}[tbh]
\includegraphics[width=0.5\textwidth]{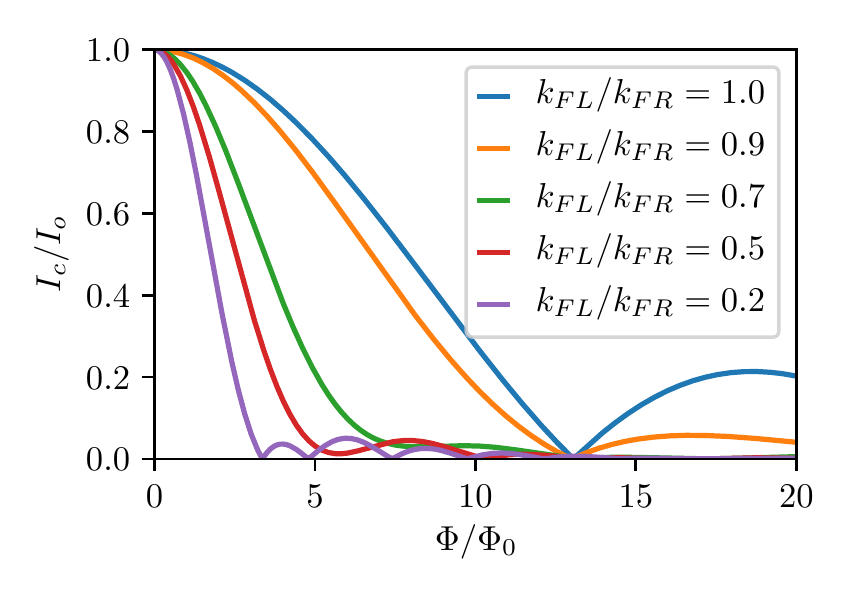}
\caption{
\label{Quasiclassicstunable}
Critical current as a function of flux $\Phi = B (W_b+W)L/2$ through the normal scattering region, calculated quasiclassically for the symmetric hourglass geometry of dimensions $L=W$ and $W_b = L/25$. The Fermi wave vector mismatch quantifies the difference of carrier densities on both sides of the hourglass bottleneck.}
\end{figure}

\comment{Considering different carrier densities on both sides of bottle-neck show a similar behaviour like asymmetry in JJ but is more feasible from an experimental point of view.}
A more practical way to break the phase matching condition of Fig.~\ref{Schematic}(c) is by tuning carrier densities across the bottleneck via a local gate potential.
We incorporate this effect in the quasiclassical calculations by introducing two Fermi wavevectors $k_{FL}$ and $k_{FR}$ on the left and right side of the bottleneck respectively.
Owing to this difference in carrier densities, a trajectory starting at $x = -L/2$ with angle $\theta$ enters the right side of the hourglass at a different angle $\theta^\prime$ which depends on the ratio of Fermi wave vectors as

\begin{equation}
\theta^\prime = \arcsin\left(\frac{k_{FL}}{k_{FR}} \sin \theta\right).
\end{equation}
As a result, the corresponding Peierls phase factor \eqref{eq:P-phase} acquires the form
\begin{equation}
\xi = \frac{2eBL}{\hbar}\left[y_0 + \left(3  \tan\theta + \tan \theta^\prime\right) \frac{L}{8}\right].
\end{equation}
(Here and later we assume $k_{FL} < k_{FR}$.)
The conditions on the angle $\theta$ of the incident trajectories for integration is given by:

\begin{subequations}
  \label{eq:theta-fermi-mismatch}
  \begin{align}
    &\arctan\left(\frac{-W_b-2y_0}{L}\right) < \theta < \arctan\left(\frac{W_b-2y_0}{L}\right),\\
    &|y_0 + L\left(\tan\theta+\tan\theta^\prime\right)/2|<W/2.
  \end{align}
\end{subequations}
Depending upon the Fermi wavevector mismatch, more trajectories can now reach the other interface without edge scattering as compared to the case of a symmetric hourglass device with equal carrier concentrations.
We show the results for different Fermi wavevector mismatch in Fig.~\ref{Quasiclassicstunable}.
Similar to making the junction itself asymmetric, introducing a carrier density mismatch restores the sensitivity of the supercurrent to the magnetic field.

\section{Tight-binding numerical calculation of supercurrent}
\label{sec:scattering-formalism}

\comment{We describe the scattering matrix formalism to calculate supercurrent using numerical simulation.}
To compare the results of the quasiclassical analysis with a quantum mechanical model, we numerically calculate the supercurrent based on a tight-binding model using the scattering matrix approach \cite{beenakker_universal_1991}.
The numerical calculations take into account effects that we neglected in quasiclassics: reflections from sample boundaries, finite Fermi wavelength, finite cyclotron radius, and potentially disorder scattering.

In the short junction limit, the scattering matrix condition for Andreev bound state reads \cite{van_heck_single_2014}:
\begin{align}
\label{eq:bound-state}
\left[ {\begin{array}{cc}
0 & -iA^{\dagger} \\
iA & 0 \\
\end{array} } \right]\Psi_\textrm{in}=E/\Delta \Psi_\textrm{in},\\
A \equiv \frac{1}{2}\left(r_A s -s^T r_A\right) \label{eq:A_definition}
\end{align}
with $\Psi_\textrm{in} = (\Psi_\textrm{in}^e, \Psi_\textrm{in}^h)$ a vector of complex coefficients describing a wave incident on the junction in the basis of modes incoming from the superconducting leads into the normal region.
The scattering matrix $s$ is due to the normal scattering region, whereas $r_A$ is due to Andreev reflection at the superconductor-normal metal interface.
In the basis where the outgoing modes are time-reversed partners of the incoming modes, the matrix $r_A$ is given by
\begin{equation}
  \label{eq:r_A_definition}
r_A = \left[ {\begin{array}{cc}
i e^{i\phi/2} \bm{1}_{n_1} & 0 \\
0 & i e^{-i \phi/2} \bm{1}_{n_2} \\
\end{array} } \right],
\end{equation}
with $\phi$ the superconducting phase difference between the two superconducting leads.

\comment{we use first order perturbation to get an expression for the phase derivative of Andreev bound state energy, thus avoiding numerical differentiation to get supercurrent.}
We square Eq.~\eqref{eq:bound-state}, making it block-diagonal and take one of the subblocks to obtain an equivalent eigenproblem for the Andreev bound states:
\begin{equation}
  \label{eq:andreev_eigenvectors}
  A^\dagger A \Psi^e_\textrm{in} = \frac{E^2}{\Delta^2} \Psi^e_\textrm{in}.
\end{equation}
Differentiating this with respect to $\phi$ we obtain:
\begin{equation}
  \label{eq:current_of_one_wf}
\frac{dE}{d\phi}=\frac{\Delta^2}{2} \frac{1}{E}\left\langle \Psi_\textrm{in}^e \left| \frac{d(A^{\dagger}A)}{d\phi}\right|\Psi_\textrm{in}^e\right\rangle.
\end{equation}
Further substituting $d (A^\dagger A) / d \phi$ from Eqs.~(\ref{eq:A_definition},\ref{eq:r_A_definition}) provides us with a closed form expresion for the supercurrent when combined with the eigenvectors from the Eq.~\eqref{eq:andreev_eigenvectors}.
We finally arrive to the supercurrent
\begin{equation}
I = -\frac{2e}{\hbar}\sum_p \tanh(E_p/2k_BT)\frac{dE_p}{d\phi},
\label{eq:Nsupercurrent}
\end{equation}
with $dE_p/d\phi$ obtained from Eqs.~(\ref{eq:andreev_eigenvectors}, \ref{eq:current_of_one_wf}).

\begin{figure}[tbh]
\includegraphics[width=0.5\textwidth]{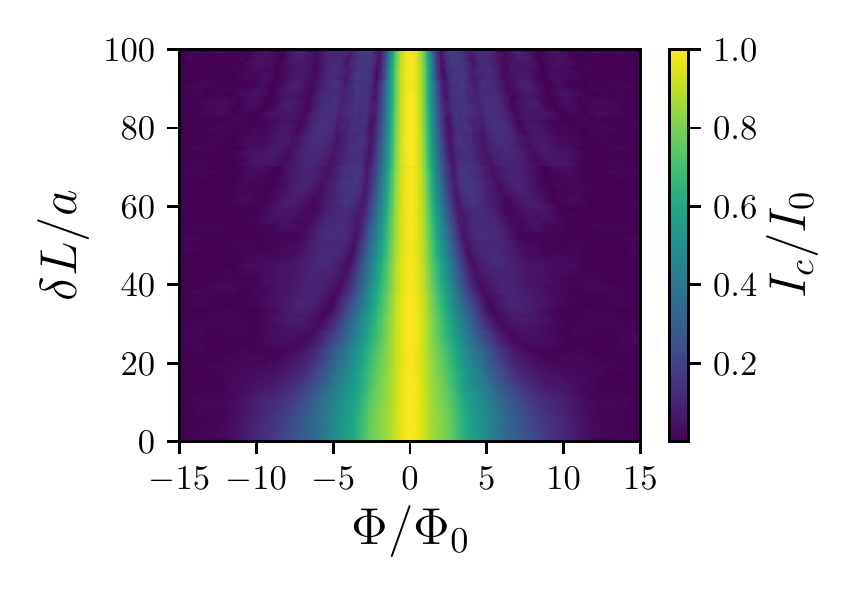}
\caption{
\label{asymmetric-numerics}
Critical current as a function of applied magnetic field and the  asymmetry of the device with $W=L=500a$ and $W_b=20a$.}
\end{figure}

\begin{figure}[tbh]
\includegraphics[width=0.5\textwidth]{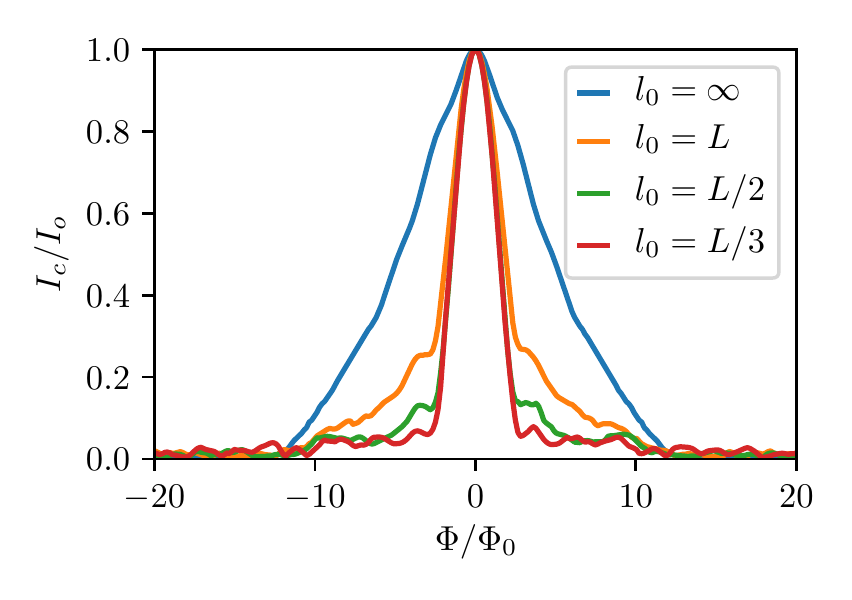}
\caption{
\label{NumericsFP}
Critical current as a function of magnetic flux through the normal scattering region for ballistic and diffusive Josephson junctions of dimensions $L=W=500a$ and $W_b=20a$, calculated from Eq.~\ref{eq:Nsupercurrent} using the tight-binding calculations. The mean free path $l_0$ is determined by the strength of disorder.}
\end{figure}

\comment{we introduce uniform disorder in our system and calculate scattering matrices using Kwant}
We calculate the normal state scattering matrix using the Kwant software package~\cite{groth_kwant:_2014}: we discretize the Hamiltonian Eq.~\eqref{eq:H-2DEG} on a square lattice with lattice constant $a$ and a shape of an hourglass, as shown in Fig.~\ref{Schematic}(b).
To analyse the effect of disorder we consider a random onsite potential, uniformly varying between $-U/2$ to $U/2$.
The quasiparticle mean free path $l_0$ in the scattering region is then given by~\cite{ando_quantum_1991}
\begin{equation}
l_0 = \frac{6 \lambda_F^3}{\pi^3 a^2} \left(\frac{\mu}{U}\right)^2,
\label{mfp}
\end{equation}
with $\lambda_F$ Fermi wavelength.
In our simulation, we chose $\mu=1.01t$, with $t$ the nearest-neighbour hopping constant.
We then evaluate the supercurrent at $T=0$.

\begin{figure*}[tb]
\includegraphics[width=0.9\textwidth]{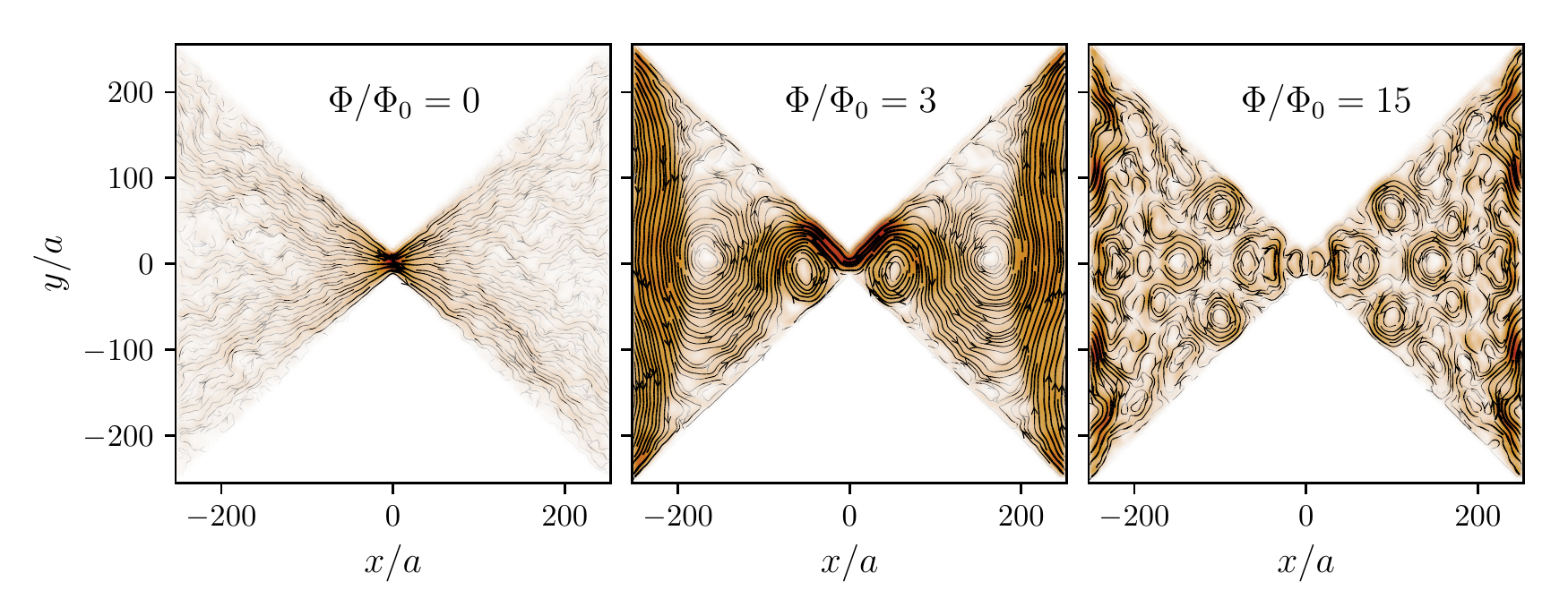}
\caption{
\label{density-maps}
Supercurrent density maps for a system of size $W=L=500a$, $W_b=20a$ at the superconducting phase $\phi = \pi/2$ for different values of total flux $\Phi = B (W_b+W)L/2$ through the normal scattering region. Left panel: At zero magnetic flux, straight trajectories give maximum supercurrent. Middle panel: At $\Phi = 3\Phi_0$, a supercurrent vortex appears, accompanies by only a slight decrease in net supercurrent [\emph{cf.} Fig.~\ref{NumericsFP}]. Right panel: At high magnetic flux $\Phi=15\Phi_0$, several supercurrent vortex appear while the net supercurrent vanishes.}
\end{figure*}

\comment{We confirm that tight-binding gives a result that is qualitatively similar to semiclassic}
To confirm the quasiclassical simulations we compute the supercurrent in an asymmetric device, with the results shown in Fig.~\ref{asymmetric-numerics} for a system of length $L=500a$, lead width $W=500a$, and the width of bottleneck $W_b = 20a$.
In a symmetric device we observe a monotonically decaying bell-shaped pattern, with the lack of the secondary lobes likely due to the small ratio $\lambda_F/W_b \approx 3$.
We observe that the predictions of the quasiclassical calculations agree with those of the fully quantum-mechanical one and confirm that the device asymmetry controls the sensitivity of the critical current to the magnetic field.

\comment{Disorder suppresses the focusing when mfp is comparable to device size.}
The effect of the disorder scattering on the geometric focusing is shown in the Fig.~\ref{NumericsFP}.
The central lobe of the Fraunhofer pattern decays much faster in the presence of a uniform disorder as compared to the ballistic case, recovering the magnetic field sensitivity of a conventional junction when $l_0 \sim L$.
This qualitative change in the Fraunhofer pattern makes the hourglass SNS junction uniquely sensitive to disorder scattering and even allows to distinguish purely ballistic transport from even quasi-ballistic transport when the mean free path is comparable to the system size.

\comment{Supercurrent density profile shows the path of current flow and the corresponding net current.}
Finally we compute the supercurrent density, as shown in Fig.~\ref{density-maps}, for three different values of magnetic flux through the device and with $\phi=\pi/2$.
The left panel shows the supercurrent distribution for no magnetic field, with the current density approximately matching that of the normal current.
In the middle panel at magnetic flux $\Phi=3\Phi_0$, we see the effect of the magnetic field which bends different trajectories in a vortex-like structure.
While the decrease of critical current at this flux value in  Fig.~\ref{NumericsFP} from zero magnetic field case is small, we see a completely different supercurrent density profile.
The additional supercurrent is mediated by the trajectories that start and end the same superconductor: in a device with a thin bottleneck, these trajectories comprise a majority.
The observation of the change in the supercurrent distribution by a scanning magnetometer~\cite{embon_imaging_2017, kirtley_response_2016} may then serve as an independent confirmation of the focusing effect.
The right panel shows a supercurrent density map at a higher magnetic field with many supercurrent vortices~\cite{cuevas_magnetic_2007, bergeret_vortex_2008, ostroukh_two-dimensional_2016} and vanishing overall supercurrent.
\section{Conclusions}
\label{sec:discussion}

\comment{Owing to the special shape of our device, there is a phase matching condition for trajectories which gives a way to probe them. An asymmetry in the device or a uniform disorder breaks this condition.}
We have proposed a strategy to observe supercurrent carried by ballistic trajectories by identifying a geometry where ballistic supercurrent vanishes at a larger magnetic field scale $\Phi_0 / W_b L$ instead of the conventional $\Phi_0/WL$.
We confirm our predictions using both quasiclassical and fully quantum-mechanical analysis and confirm that breaking the phase cancellation condition leads to a faster decay of the central lobe and a conventional Fraunhofer pattern.
Although we consider a conventional two-dimensional electron gas in our analysis, we expect that the proposed phenomenon should exist in any mesoscopic Josephson device due to being a geometrical effect.
Therefore, the proposed device design is well within the reach of the current experimental technology and can be implemented using both semiconducting quantum wells~\cite{suominen_anomalous_2017} or high quality graphene Josephson junctions~\cite{allen_spatially_2016, kraft_tailoring_2018}.

The source code and data used for figures in this work is available at~\onlinecite{irfan_geometric_2018}.
 
\acknowledgements{
The authors thank D. Sticlet, M. P. Nowak, and B. Nijholt for useful discussions. This work was supported by the Netherlands Organization for Scientific Research (NWO/OCW), as part of the Frontiers of Nanoscience program and an European Research Council (ERC) Starting Grant.}

\emph{Author contributions:}
A. Akhmerov proposed the idea and supervised the project.
M. Irfan performed the analytical and numerical calculations.
Both authors contributed to writing the manuscript.

\bibliographystyle{apsrev4-1}
\bibliography{hourglass}

\end{document}